\documentclass[prl,twocolumn,showpacs]{revtex4-1}

\usepackage{graphicx}
\usepackage{amsmath}

\newcommand{\cs}{c_{\rm s}}
\newcommand{\Dbar}{\bar{D}_{\rm col}}
\newcommand{\Dc}{D_{\rm c}}
\newcommand{\Dcol}{D_{\rm col}}
\newcommand{\Ds}{D_{\rm s}}
\newcommand{\Dself}{D_{\rm self}}
\newcommand{\eg} {{e.g., }}
\newcommand{\etabar}{\bar{\eta}}

\newcommand{\ie} {{i.e., }}
\newcommand{\kT} {k_{\rm B}T}

\begin{document}

%==============================================================
\title{Long-Range Dynamic Correlations in Confined Suspensions} 
%==============================================================

\author{Derek Frydel}
\affiliation{Raymond \& Beverly Sackler School of Chemistry, Tel Aviv
University, Tel Aviv 69978, Israel}

\author{Haim Diamant} 
\email{hdiamant@tau.ac.il} 
\affiliation{Raymond \& Beverly Sackler School of Chemistry, Tel Aviv
University, Tel Aviv 69978, Israel}

\date{June 17, 2010}

\begin{abstract}
  Hydrodynamic interactions between particles confined in a
  liquid-filled linear channel are known to be screened beyond a
  distance comparable to the channel width. Using a simple analytical
  theory and lattice-Boltzmann simulations, we show that the
  hydrodynamic screening is qualitatively modified when the
  time-dependent response and finite compressibility of the host
  liquid are taken into account. Diffusive compression modes in the
  confined liquid cause the particles to have velocity correlations of
  unbounded range, whose amplitude decays with time only as
  $t^{-3/2}$. 
%In the case of a ring-like channel a uniform
%  (zero-wavenumber) compression mode causes long-range correlations at
%  steady state as well, thus strongly affecting the collective
%  mobility of the particle assembly.
\end{abstract}


\maketitle
%------------------------------------------------

As particles move through a liquid, the flows that they create
correlate their motions. These hydrodynamic interactions play a
central role in the dynamics of particulate liquids \cite{colloids}.
In unbounded liquids the correlation is dominantly mediated by
diffusive shear modes (\ie diffusion of transverse momentum) as
accounted for by the Navier-Stokes equation for an incompressible
liquid. Consequently, the perturbation caused by the forced motion of
a particle in the liquid can be represented at large distances as
emanating from a momentum monopole.  The resulting flow velocity field
decays with distance as $1/r$, leading to long-range hydrodynamic
interaction between particles and the well studied phenomenon of long
time tails in their velocity correlation functions \cite{Hansen}.

When the suspension is confined by rigid boundaries, the hydrodynamic
interaction is qualitatively different \cite{jpsj09}. Because of
momentum transport into the boundaries the liquid no longer conserves
momentum, and the amplitude of shear modes, developing in the liquid
in response to particle motion, decays exponentially with distance.
The correlations are now dominantly mediated by compression modes, and
the perturbation due to particle motion can be represented at large
distances as arising from an effective mass dipole.  When the
suspension is confined between two plates (quasi-two-dimensional
suspension), the resulting flow field remains long-ranged, decaying as
$1/r^2$ (as for a mass dipole in two dimensions) \cite{prl04}.
However, since in one dimension (1D) a steady dipole has a vanishing
effect, the steady flow due to local particle motion in a linear
channel has no far field and decays exponentially with $r/h$, $h$
being the channel width \cite{LironShahar,prl02}.  This hydrodynamic
screening in pores and channels is well documented (\eg
\cite{prl02,Beatus}) and widely employed (for instance, in studying
the dynamics of confined polymers \cite{deGennes}).

The analysis above assumes that the host liquid is at steady state and
is incompressible, thus omitting longitudinal sound modes. Since
the Reynolds number of relevant colloidal systems is typically smaller
than $10^{-6}$, and the sound velocity in the liquid, $\cs$, is of
order $10^3$ m/s, both assumptions seem safely valid.  Yet, as was
first pointed out in Ref.\ \cite{Frenkel} based on simulations, and
further established analytically \cite{Felderhof,Derek}, confinement
by rigid boundaries strongly affects the sound modes in a compressible
liquid.  These modes change from underdamped propagation with velocity
$\cs$ to overdamped diffusion with diffusivity $\Ds\sim\cs^2 h^2/\nu$,
where $\nu$ is the kinematic viscosity of the liquid. The resulting
sound diffusion is fast, $\Ds\sim 1$ m$^2$/s for water in micron-scale
confinement, compared to $\nu\sim 10^{-6}$ m$^2$/s for the diffusion
coefficient of shear modes, yet the latter are suppressed by the
boundaries, leaving the diffusive compression modes as the sole
mechanism for hydrodynamic interaction at interparticle distances
$r>h$.  Furthermore, similar to the appearance of long time tails due
to shear-stress diffusion in unbounded suspensions \cite{Hansen}, the
diffusive nature of compression modes in confined liquids leads to a
(negative) long time tail in the velocity autocorrelation function of
a suspended particle \cite{Frenkel}.
In this Letter we show that those diffusive compression modes have far
reaching implications for the dynamic correlations between distant
particles in a channel.

We employ a simple phenomenological approach, in which only the
unconfined dimensions (in the current case the single dimension along
the channel) are retained, whereas all the details in the confined
dimensions, such as the cross-section of the channel and the precise
boundary conditions at its walls, are averaged over and reduced to an
effective friction term. Such a lubrication approach
\cite{lubrication} leads to tremendous simplification of the analysis
while yielding the correct physics at distances larger than the
confinement width. It was successfully applied in the past to
various problems of confined hydrodynamics (\eg \cite{Mazenko})
and, in particular, was shown to correctly reproduce the diffusive
sound modes in confinement \cite{Frenkel,Derek}.

We support the calculations by three-dimensional lattice-Boltzmann
simulations \cite{LB}. In the simulations spherical particles of
diameter $\sigma$ move along the axis of a channel of length $L$ and a
square cross-section of side $h$. The channel is filled with an
ambient fluid of mass density $\rho_0$, shear viscosity $\eta$, bulk
viscosity $\zeta$, and sound velocity $\cs$. The channel and particles
are rigid, and a no-slip boundary condition is imposed at their
surfaces using the second-order bounce-back scheme \cite{LB}.
Periodic boundary conditions are imposed at the edges of the channel,
though $L$ is taken sufficiently large
%($10^4$ max for the longest time 8000) 
to render finite-length effects negligible during the simulated
time.  In terms of lattice spacings and simulation time steps the
parameter values are $h=15$, $\sigma=10$, $\rho_0=1$, $\eta=1/6$,
$\zeta=1/9$, and $\cs=1/\sqrt{3}$. To present the simulation results
in a way relevant to real systems we shall use as units of length,
time, and diffusivity, respectively, the channel width $h$, the time
it takes shear modes to diffuse to the boundaries, $\tau\equiv
h^2/\nu$ ($\nu=\eta/\rho_0$), and the self-diffusion coefficient of an
unbounded particle, $D_0\equiv\kT/(3\pi\eta\sigma)$, $\kT$ being the
thermal energy. For particles of $\sigma=1$ $\mu$m in a water-filled
channel of $h=1.5$ $\mu$m at room temperature, we have $\tau\simeq 2$
$\mu$s and $D_0\simeq 0.4$ $\mu$m$^2$/s.

Within the simplified theory the linearized 1D hydrodynamic equations,
corresponding to a channel, read
\begin{eqnarray}
  &&\rho_0\dot{u} = -p' + (4\eta/3+\zeta)u'' - (\alpha\eta/h^2) u 
  + (\beta/h^2)f,
 \nonumber\\
  &&\dot{\rho} = -\rho_0 u',\ \ \ 
  p = \cs^2\rho.
\label{hydro}
\end{eqnarray}
Being one-dimensional, Eq.\ (\ref{hydro}) describes solely
longitudinal flow. In it a dot denotes a derivative with respect to
time $t$, and a prime is a derivative with respect to distance $x$
along the channel. The fields $\rho(x,t)$, $p(x,t)$, and $u(x,t)$ are,
respectively, the perturbations in liquid mass density, pressure, and
velocity about $\rho_0$, $p_0$, and $0$, in response to the force
density (per unit length) $f(x,t)$.
%The coefficients are the shear
%viscosity $\eta$, bulk viscosity $\zeta$, and sound velocity $\cs=(\pd
%p_0/\pd\rho_0)^{1/2}$. 
We have introduced two dimensionless coefficients --- $\alpha$,
determining the strength of the effective friction caused by the
boundaries, and $\beta$, equal to the ratio between $h^2$ and the
cross-sectional area of the channel. For the simulated square channel
$\beta=1$, and we find $\alpha\simeq 28$ by a fitting procedure
independent of the following analysis \cite{ft_alpha}. 

The velocity Green's function, $u(x,t)=G(x,t)$, obtained from Eq.\ 
(\ref{hydro}) upon substituting $f(x,t)=\delta(x)\delta(t)$, is
readily calculated in Fourier space [$G(q,\omega)=\iint dxdt
e^{-i(qx-\omega t)}G(x,t)$] as
\begin{equation}
  G(q,\omega) = \frac{(\beta/h^2)} %e^{-i(qx'-\omega t')}}
  {( \etabar + i\cs^2\rho_0/\omega )q^2 + \alpha\eta/h^2 
  - i\rho_0\omega},
\label{Green1}
\end{equation}
where $\etabar\equiv 4\eta/3+\zeta$. Inverting the spatial coordinate
back to real space, we obtain
\begin{eqnarray}
  G(x,\omega) &=& A(\omega) e^{-|x|/\lambda(\omega)} \nonumber\\
  A(\omega) &=& [\beta/(2h^2)] \left[ (\etabar + i\cs^2\rho_0/\omega)
  (\alpha\eta/h^2 - i\rho_0\omega) \right]^{-1/2} \nonumber\\
  \lambda(\omega) &=& \left( \frac{\etabar + i\cs^2\rho_0/\omega}
  {\alpha\eta/h^2 - i\rho_0\omega} \right)^{1/2}.
\label{Green2}
\end{eqnarray}
The function $G(x,t)$ gives the flow velocity in the channel as a
function of position and time in response to a point impulse applied
at the origin at $t=0$. If two particles are positioned inside the
channel at sufficiently large mutual distance, $x\gg \sigma$, the same
function, up to a dimensionless factor, $\gamma(\sigma/h)$, gives the
velocity of one particle in response to a unit momentum imparted to
the other. Hence, the velocity cross-correlation function of the
particle pair is given, according to the fluctuation-dissipation
theorem, by $C(x,t)\equiv\langle V_1(0)V_2(t)\rangle(x) = \gamma\kT
G(x,t)$ \cite{Hansen}. The coupling diffusion coefficient,
characterizing the correlated Brownian motion of the pair, is then
obtained from the appropriate Green-Kubo relation \cite{Hansen},
$\Dc(x)=\int dt C(x,t)=\gamma\kT G(x,\omega=0)$.

\begin{figure}[tbh]
\vspace{0.6cm}
\centerline{\resizebox{0.45\textwidth}{!}
{\includegraphics{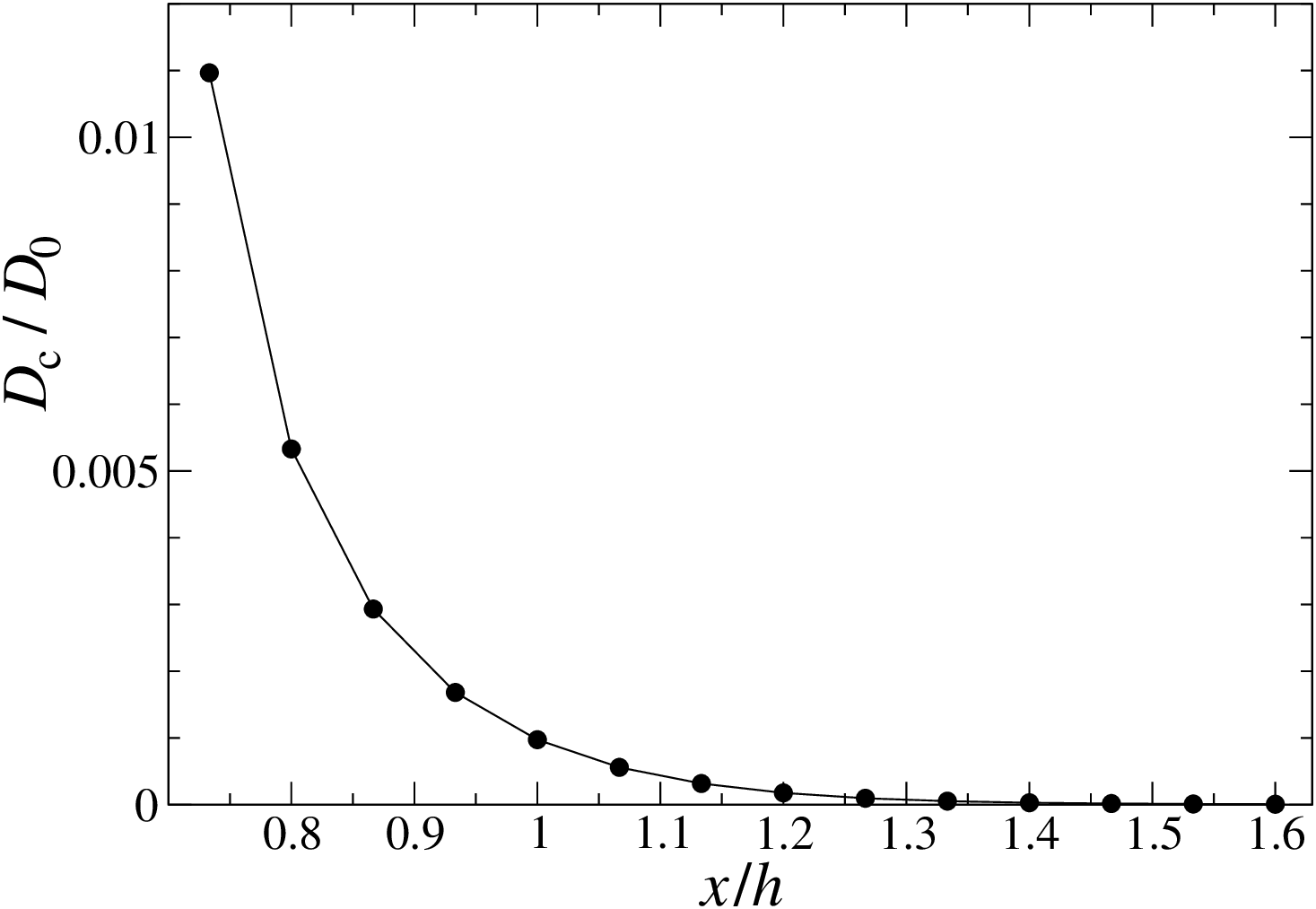}}}
\caption{Coupling diffusion coefficient of a particle pair in a channel
  as a function of interparticle distance, obtained from
  lattice-Boltzmann simulations. The coefficient is measured from the
  net displacement (time-integrated velocity) of one particle in
  response to a unit-momentum impulse imparted to the other. The
  particle diameter is $\sigma=h/1.5$.}
\label{fig_Dc}
\end{figure}

\begin{figure}[tbh]
\vspace{0.25cm}
\centerline{\resizebox{0.43\textwidth}{!}
{\includegraphics{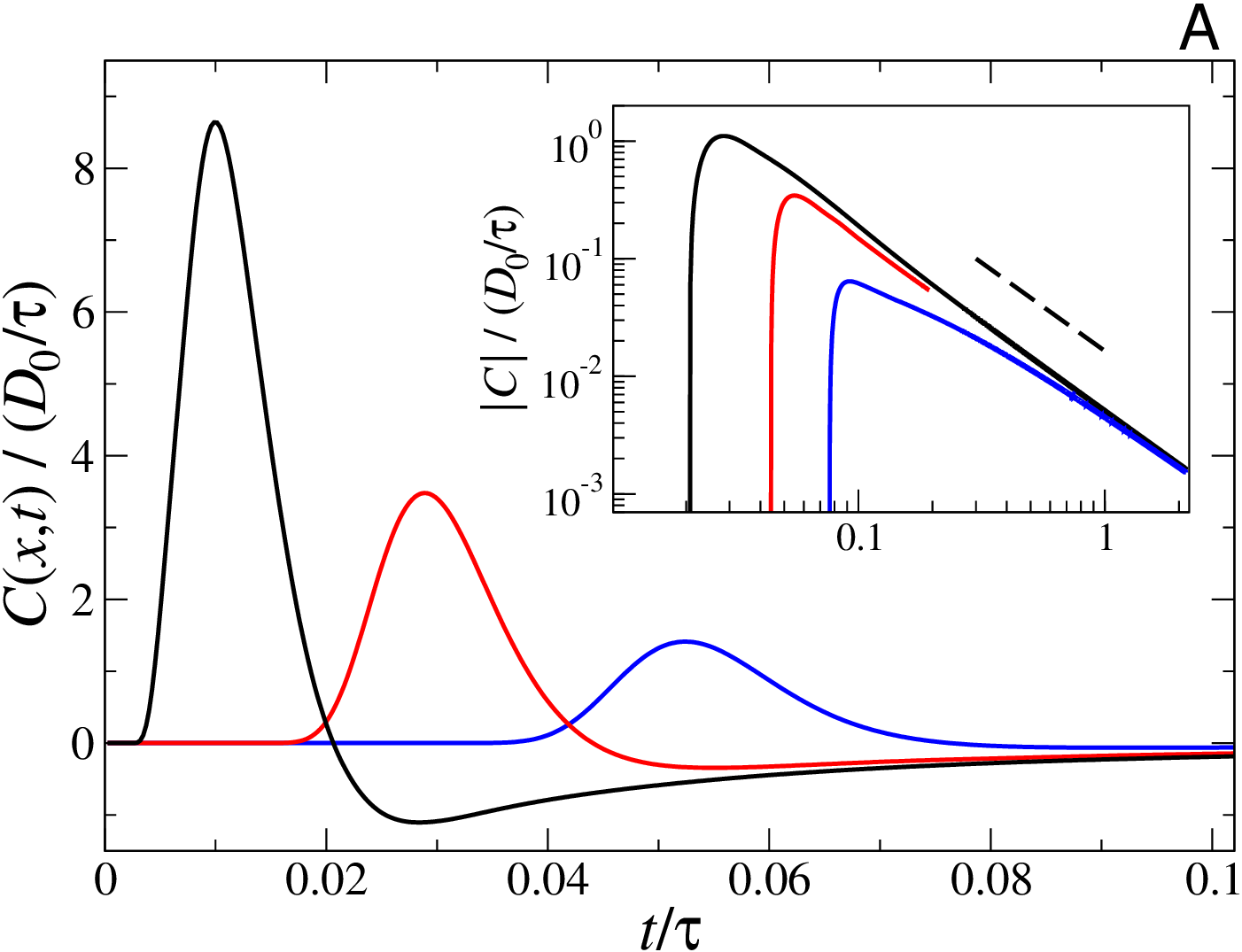}}}
\vspace{0.7cm}
\centerline{\resizebox{0.43\textwidth}{!}
{\includegraphics{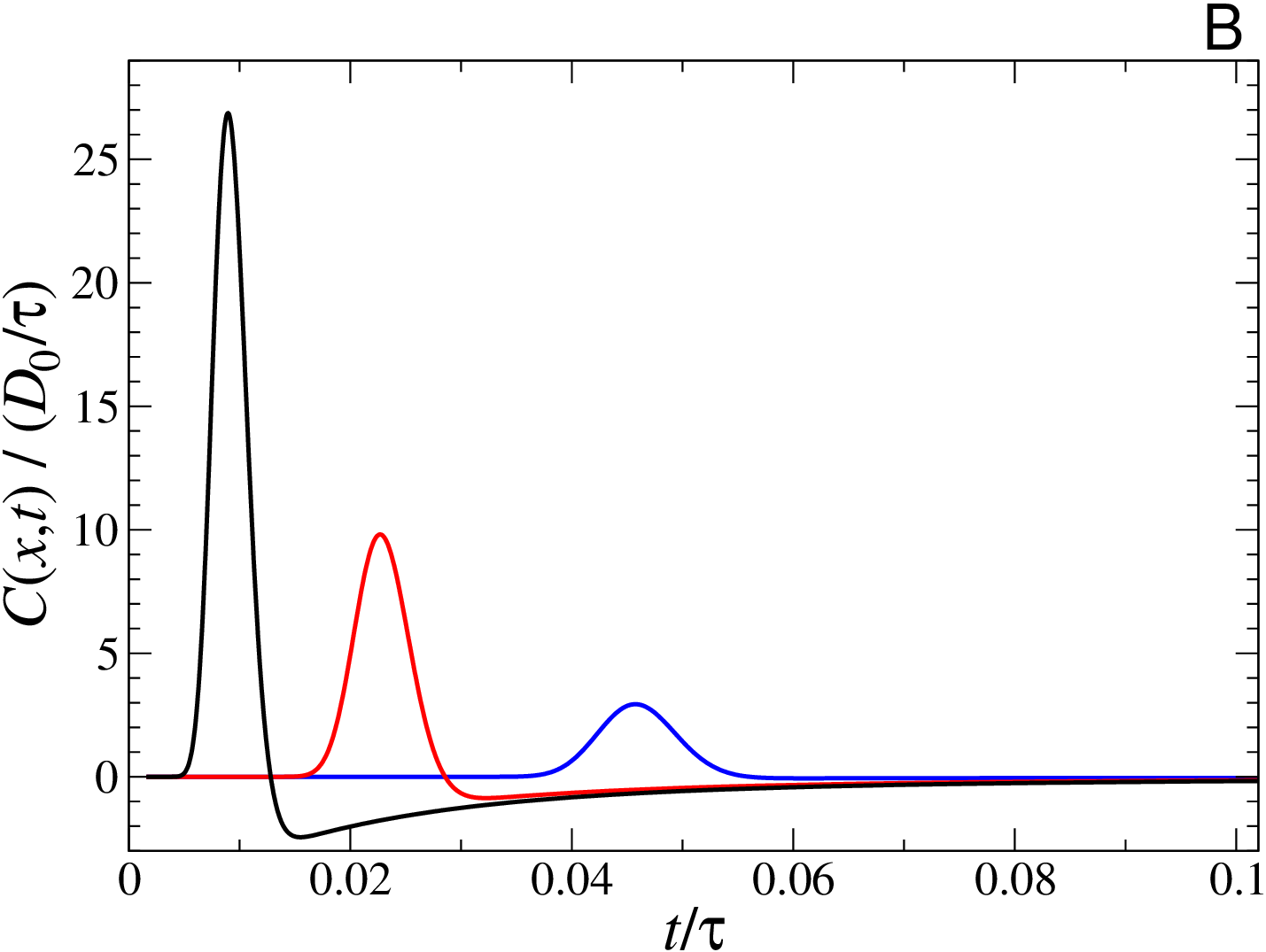}}}
\caption{(color online). Velocity cross-correlation function of a 
  particle pair in a channel as a function of time. Different curves
  correspond to different interparticle distances (from left to
  right): $x/h=1.33$ (black), $3.33$ (red), and $6.67$ (blue). The
  particle diameter is $\sigma=h/1.5$. (A) Lattice-Boltzmann
  simulation results. The correlation is measured from the
  time-dependent velocity of one particle in response to a
  unit-momentum impulse imparted to the other. The inset shows the
  long-time behavior on a log-log scale, the dashed line indicating a
  slope of $-3/2$.  (B) Results from the analytical approximation
  [inversion of Eq.\ (\ref{Green2})].}
\label{fig_corr}
\end{figure}

Two important observations readily follow from Eq.\ (\ref{Green2}).
First, since $A(\omega=0)=0$, the coupling diffusion coefficient
vanishes, $\Dc(x)=0$. This is the manifestation of steady-state
hydrodynamic screening in the channel. The actual exponential decay of
$\Dc(x)$ arises from the screened transverse modes, which have been
eliminated in the simplified 1D calculation. It is clearly seen in the
simulation results shown in Fig.\ \ref{fig_Dc}.  Note that periodic
boundary conditions have not been imposed in Eq.\ (\ref{Green2}),
allowing a pressure difference to form between the edges of the
channel \cite{LironShahar,Bhattacharya}.  Second, as the diffusive
response broadens, the correlation length increases indefinitely,
$\lambda\sim\omega^{-1/2}$ for small $\omega$, while the correlation
amplitude decreases as $A\sim\omega^{1/2}$. This leads at long times
to a correlation of unbounded range and slowly decaying negative
amplitude,
\begin{equation}
  C(x,t\rightarrow\infty)/(\gamma\kT) 
  \simeq -\frac{\beta}{4\alpha\eta(\pi\Ds)^{1/2}} t^{-3/2},
\label{tail}
\end{equation}
where $\Ds=\cs^2h^2/(\alpha\nu)$. To reach this distance-independent
behavior the diffusive front must have already passed the distance
$x$, \ie $t>x^2/\Ds$. For $x\sim 1$ cm in a micron-wide channel filled
with water this requires $t>10^{-4}$ s, \ie Eq.\ (\ref{tail}) holds
practically at all times. (For liquids of higher viscosity, however,
the distance-independent regime will take a longer time to establish.)
The negative $t^{-3/2}$ tail is confirmed by the simulation (Fig.\ 
\ref{fig_corr}A inset), from which we get for the square channel
$\gamma(\sigma/h=2/3)\simeq 2.61$ \cite{ft_gamma}.

In slightly more detail, the temporal correlation between the
velocities of two particles consists of three parts (Fig.\ 
\ref{fig_corr}). Because of the finite speed of sound there is a short
incipient period of no correlation, which becomes longer with
increasing interparticle distance. It is followed by a short peak of
positive correlation upon the arrival of the signal and a subsequent
long negative tail, which is the result of sound diffusion through the
channel.  As can be seen in Fig.\ \ref{fig_corr}B, the simple theory
qualitatively captures all three features. As the interparticle
distance becomes larger than $h$, the quantitative agreement between
theory and simulation progressively improves.

At steady state ($\omega=0$) the correlation becomes exponentially
small in $x/h$ (Fig.  \ref{fig_Dc}). This is because the long-ranged
temporal correlation arises from longitudinal modes, and these cannot
produce a net displacement for any finite wavelength. (If they could,
it would imply a nonuniform steady density of the liquid.)  Thus, the
areas under the theoretical curves of Fig.\ \ref{fig_corr}B all
vanish, and the remaining steady-state coupling is produced solely by
the screened shear modes (the finite total areas under the simulated
curves of Fig.\ \ref{fig_corr}A).

\begin{figure}[tbh]
  \vspace{0.65cm} \centerline{\resizebox{0.42\textwidth}{!}
    {\includegraphics{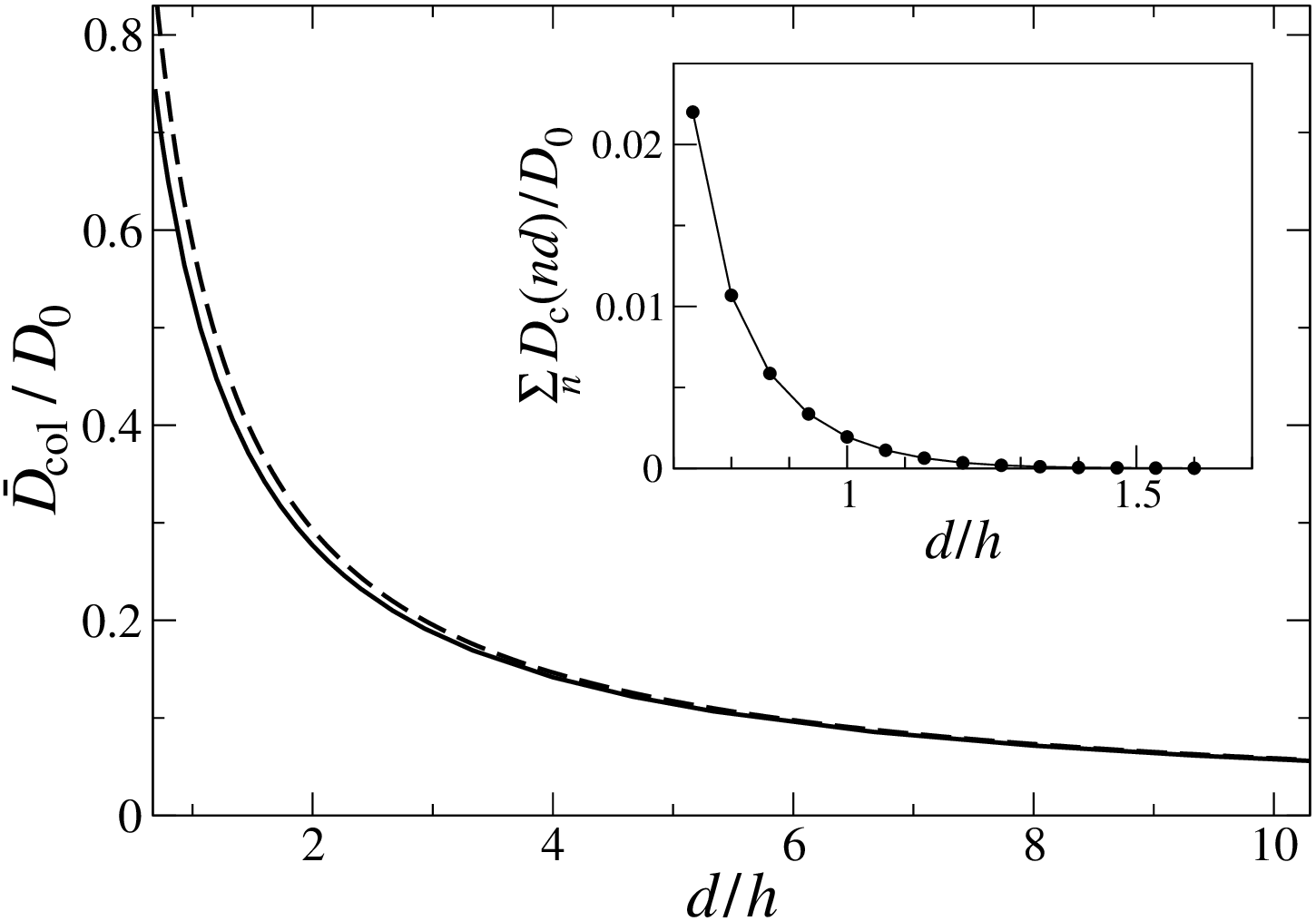}}}
\caption{Collective diffusion coefficient of an array of particles
  as a function of interparticle separation.  Lattice-Boltzmann
  simulation results (solid) are presented alongside the analytical
  result [Eq.\ (\ref{Dcol}); dashed]. The inset shows the collective
  coefficient as expected from the hydrodynamic-screening description.
  The particle diameter is $\sigma=h/1.5$.}
\label{fig_Dcol}
\end{figure}

So far the way to make the unscreened longitudinal modes effective has
been to include liquid compressibility. Another specific but
noteworthy example of an unscreened response is due to the uniform
($q=0$) mode, which is unaffected by the incompressibility constraint.
Indeed, from Eq.\ (\ref{Green1}) one has $G(q\neq 0,\omega=0)=0$ but
$G(q=0,\omega=0)=\beta/(\alpha\eta)>0$.  Consider an infinite array of
particles along the channel, separated by a distance $d$ from each
other, and examine the collective diffusion coefficient of the array,
$\Dcol$, corresponding to collective motion of all the particles. For
$d\rightarrow\infty$ $\Dcol=\Dself$, the self-diffusion coefficient of
an isolated particle in the channel.  Thus,
$\Dbar\equiv\Dcol-\Dself=2\sum_{n=1}^\infty\Dc(x=nd)$ characterizes
the contribution to the collective motion from flow-induced
correlations.  With only transverse modes in mind, since each of the
pair couplings $\Dc(x)$ is screened (Fig. \ref{fig_Dc}), we should
expect $\Dbar$ to decay exponentially with $d/h$. This behavior,
expected from conventional hydrodynamic screening, is depicted in the
inset of Fig.\ \ref{fig_Dcol}. However, considering the longitudinal
flow of Eq.\ (\ref{Green2}), we get
\begin{equation}
  \Dbar/(\gamma\kT) = 2\sum_{n=1}^\infty G(x=nd,\omega) 
  |_{\omega\rightarrow 0} = \beta / (\alpha\eta d),
\label{Dcol}
\end{equation}
\ie a much slower decay of $\Dbar\sim 1/d$. The appearance of $1/d$ in
$\Dcol$, in place of the $1/\sigma$ in $D_0$, implies that the
particles, no matter how far apart, carry the liquid column in between
them as they move collectively through the channel. This again
reflects an interparticle coupling of unbounded range.
As suggested by the absence of $\rho_0$ and $\cs$ in Eq.\ 
(\ref{Dcol}), this result is also obtainable from a simpler
incompressible theory. In the $q=0$ case the incompressibility
constraint, as well as the periodic boundary conditions, are trivially
obeyed.  There is no pressure difference between the edges of the
channel and, consequently, each forced particle creates a uniform flow
velocity $\sim(\eta L)^{-1}$ \cite{Bhattacharya}, which affects all
others. Thus, the collective mobility of $N$ particles is $\sim N(\eta
L)^{-1}=(\eta d)^{-1}$, in accord with Eq.\ (\ref{Dcol}). It follows
that the same result is valid for a finite ring-like channel and for a
disordered train of particles, whereby the factor $1/d$ in Eq.\ 
(\ref{Dcol}) is replaced with the mean linear density of particles.

The collective mode of an infinite particle array is easily simulated
by considering a single particle at the center of a channel of length
$d$ and imposing periodic boundary conditions in the $x$ direction.
The collective mobility is then obtained from the net displacement
(time-integrated velocity) of the particle in response to an initial
impulse. As shown in Fig.\ \ref{fig_Dcol}, the simulation results are
in good agreement with Eq.\ (\ref{Dcol}), exhibiting a slow $1/d$
decay. (No fitting parameters are used in Fig.\ \ref{fig_Dcol} in
addition to the already extracted values of $\alpha\simeq 28$ and
$\gamma\simeq 2.61$.)

In summary, the common view, according to which particles move through
liquid-filled channels and pores in an essentially uncorrelated way,
is quite misleading. Before steady state is reached, the motions of
confined particles are in fact correlated over large distances and
long times.  The physical mechanism behind the correlations is
fundamentally different from that in the unconfined case, as it
originates from the confinement-affected longitudinal (rather than
transverse) liquid response \cite{ft_exponent}.  Consequently, the
long-ranged correlations are negative (rather than positive) and do
not give rise to correspondingly long-ranged steady correlations, so
that at steady state hydrodynamic screening is recovered.  These
long-ranged temporal correlations may have important implications, for
example, for suspended particles in porous media or the dynamics of
polymer translocation through narrow pores.
Despite the small factor $\Ds^{-1/2}\sim\cs^{-1}$ appearing in Eq.\ 
(\ref{tail}), they should be observable when probing sufficiently
short times. For instance, over a millisecond (kHz frequency) Eq.\ 
(\ref{tail}) predicts for particles in a micron-wide channel filled
with water a velocity cross-correlation of order $-10^{-2}$
($\mu$m/s)$^2$, {\em independent of particle separation}.
In addition, the predicted strong effect that arises from the uniform
longitudinal mode [Eq.\ (\ref{Dcol})] can be readily checked by
measuring the collective mobility or diffusivity of a dilute particle
assembly confined in a closed ring-like microfluidic channel.
%

%------------------------------------------------
\begin{acknowledgments}
  We thank E.\ Rabani for help in the simulations and T.\ Beatus, S.\ 
  Bhattacharya, J.\ Blawzdziewicz, S.\ Rice, and T.\ Tlusty for
  helpful discussions.  This research has been supported by the Israel
  Science Foundation (ISF) under Grant No.\ 588/06. DF acknowledges
  partial support from the Israel Council for Higher Education.
\end{acknowledgments}

%------------------------------------------------
% References
%------------------------------------------------

%------------------------------------------------


\begin{thebibliography}{99}
\bibitem{ft_alpha} Equation (\ref{hydro}) can be used also to
  calculate $\rho(x,t)$ in response to an initial local density
  perturbation, $\rho(x,0)=\rho_0+\rho_1\delta(x)$. At long time the
  density field is a Gaussian with $\langle x^2\rangle=2\Ds t$,
  $\Ds=\cs^2h^2/(\alpha\nu)$. We fit $\alpha$ by comparison with a
  simulation of such a perturbation.
\bibitem{ft_gamma} In the simulation the velocity is measured on the
  channel axis, whereas in the 1D theory it represents a
  cross-sectional average; this affects the fitted value of $\gamma$.
\bibitem{ft_exponent} The exponent $-3/2$ in the velocity correlation
  function being identical to that in an unconfined liquid is a sheer
  coincidence.  In the unconfined case it is equal to
  $-D/2$ with dimensionality $D=3$, whereas in the channel it is equal
  to $-(D/2+1)$ with $D=1$.
\end{thebibliography}
\end{document}